\documentclass[twocolumn,showpacs,preprintnumbers,amsmath,amssymb,jap]{revtex4-1}

\usepackage{graphicx}
\usepackage{dcolumn}
\usepackage{bm}
\usepackage{color}
\usepackage{multirow}

\begin{document}

\title{Indium segregation during III-V quantum wire and quantum dot formation on patterned substrates}

\author{Stefano T. Moroni$^1$, Valeria Dimastrodonato$^1$, Tung-Hsun Chung$^1$, Gediminas Juska$^1$, Agnieszka Gocalinska$^1$, Dimitri D. Vvedensky$^2$, and Emanuele Pelucchi$^1$}

\affiliation{$^1$Tyndall National Institute, ``Lee Maltings'', University College Cork, Cork, Ireland,}

\affiliation{$^2$The Blackett Laboratory, Imperial College London, London SW7 2AZ, United Kingdom}

\begin{abstract}
We report a model for metalorganic vapor-phase epitaxy on non-planar substrates, specifically V-grooves and pyramidal recesses, which we apply to the growth of InGaAs nanostructures.  This model, based on a set of coupled reaction-diffusion equations, one for each facet in the system, accounts for the facet-dependence of all kinetic processes (e.g.~precursor decomposition, adatom diffusion, and adatom lifetimes), has been previously applied to account for the temperature-, concentration-, and temporal-dependence of AlGaAs nanostructures on GaAs (111)B surfaces with V-grooves and pyramidal recesses. In the present study, the growth of In$_{0.12}$Ga$_{0.88}$As quantum wires at the bottom of V-grooves is used to determine a set of optimized kinetic parameters. Based on these parameters, we have modeled the growth of In$_{0.25}$Ga$_{0.75}$As nanostructures formed in pyramidal site-controlled quantum-dot systems, successfully producing a qualitative explanation for the temperature-dependence of their optical properties, which have been reported in previous studies. Finally, we present scanning electron and cross-sectional atomic force microscopy images which show previously unreported facetting at the bottom of the pyramidal recesses that allow quantum dot formation.
\end{abstract}

\maketitle

\section{Introduction}

Tuning the electrical and optical properties of advanced III-V quantum-effect-based nanostructures, while controlling their position on a chip is crucial for quantum optic\cite{kok07,politi09} and optoelectronic\cite{noda06,atlasov11} applications. For example, coupling their optical emission to optical cavities or photonic crystal waveguides requires precise spectral and positional control\cite{hennessy07}.  Of the available growth techniques, epitaxy on patterned substrates exploits the different precursors and adatom kinetics on different facets, which influences local growth rates and local compositions, depending, e.g., on the substrate temperature, V/III ratio and overall deposition rate. Hence, as well as the seeding of the nanostructures, the patterning also allows control over their dimensions and, consequently, over their optical properties\cite{kapon04,kiravittaya09}.

Metal-organic vapor-phase epitaxy (MOVPE) on patterned substrates of V-grooves quantum wires and pyramidal quantum dots\cite{martinet98,hartmann97} has made important contributions to this field in the last 20 years, because of the precise control over the dimensions and position of the nanostructures, together with high degree of uniformity of emission properties\cite{mereni09,pelucchi03}. A noteworthy advantage of this approach is the fabrication of arrays of devices, as recently demonstrated in Ref.~\onlinecite{juska13}, where entangled photon emission from an array of In$_{0.25}$Ga$_{0.75}$As nanostructures was reported.

Some of us have recently presented a phenomenological model for MOVPE growth of (Al)GaAs on V-grooved substrates and pyramidal recesses\cite{pelucchi11}.  The model is expressed as coupled rate equations, one for each facet, and takes into account the interplay between the precursor decomposition rate, adatom diffusion and incorporation, all of which are facet-dependent processes.  By comparing with systematic experiments, this model produces quantitative agreement with the observed morphological evolution of the surfaces and the compositional dependence on position for both transient and stationary growth regimes as a function of temperature\cite{dimas12,dimas13,dimas14}.

In this work we extend the model to the simulation of the MOVPE of InGaAs, which is extensively employed as optically active layer in quantum dots and quantum wires. As a first step, the growth of In$_{0.12}$Ga$_{0.88}$As quantum wires is studied to determine a set of optimized kinetic parameters for InGaAs epitaxy that, when used in our model, reproduce the experimental growth evolution. These optimized parameters are then used to model the growth of In$_{0.25}$Ga$_{0.75}$As for nanostructures formed in the pyramidal quantum dot (QD) system. These include an In$_{0.25}$Ga$_{0.75}$As quantum dot layer sandwiched in GaAs barrier layers at the bottom of the pyramidal recess and three lateral quantum wires (LQWRs) along its edges\cite{dimas10,hartmann99}. We find that our model can qualitatively explain an unexpected experimental evidence reported in previous studies (see Ref.~\onlinecite{juska14}) where, by diminishing the growth temperature, a blue-shift of the QD emission was observed, while the LQWRs surprisingly showed the opposite trend. 

The facetting of the surfaces composing the profile of the pyramidal recess plays a major role in the evolution of the nanostructures grown on it. Several studies report the formation of high-index facets during MOVPE growth of III-V nanostructures\cite{martinet98,yuan09}, which can considerably affect the growth result. In the final section we show experimental microscopy data showing that InGaAs QD formation in pyramidal recesses is accompanied by a  more complex than expected facetting.  We conclude with discussion of open issues.

\section{Theory}

Our growth model takes into account the following processes, which, in a simplified picture, are assumed to determine the main aspects of growth by  MOVPE.  Precursors (trimethylgallium/aluminum/indium as group-III and arsine as group-V atom sources) arrive on the surface of the substrate and, after diffusing, decompose, releasing single atoms of the growing material while the remaining reactants desorb from the surface.  The released atoms then diffuse on the surface until incorporation into the growth front. The high V/III precursor flows ratio employed experimentally ($\sim$600) enable us to consider the kinetics of only the group-III species and neglect the kinetics associated with the group-V species, as they are unlikely to be a rate-limiting. Analogous assumptions are made for modelling molecular-beam epitaxy of III-V systems\cite{shitara92}. For each of the group-III species comprising the alloy, the evolution of the free-atom density $n_i$ on each facet $(i)$ can be determined through the reaction-diffusion equation:
\begin{equation}
{\partial n_i\over\partial t}=D_i\bm{\nabla}^2n_i+F_i-{n_i\over\tau_i}\, ,
\label{eq1}
\end{equation}
where $D_i$ is the diffusion coefficient, $F_i$ is the effective single atom deposition rate (which is affected by the anisotropy of the decomposition rate of the precursors), and $\tau_i$ is the average adatom lifetime prior to incorporation. The diffusion coefficient and adatom lifetime are taken to have Arrhenius forms:~$D_i=D_0e^{-\beta E_{D,i}}$ and $\tau_i^{-1}=\nu_0e^{-\beta E_{\tau_i}}$, in which $\beta=1/(k_BT)$, $k_B$ is Boltzmann's constant, $T$ the absolute temperature, and $E_{D,i}$ and $E_{ \tau,i}$ the energy barriers, respectively, for the diffusion and the incorporation processes.  This form emerges directly from transition-state theory\cite{hanggi90}, but we treat the Arrhenius parameters (prefactors and barriers) as adjustable.  We have used $D_0=a^2\nu$, where $a$ is the lattice constant of the surface, with $\nu=10^{15}$\,Hz,
while $\nu_0=4.59$~Hz (Ref.~\onlinecite{dimas13}).

\begin{figure}[t]
\includegraphics[width=8.3cm]{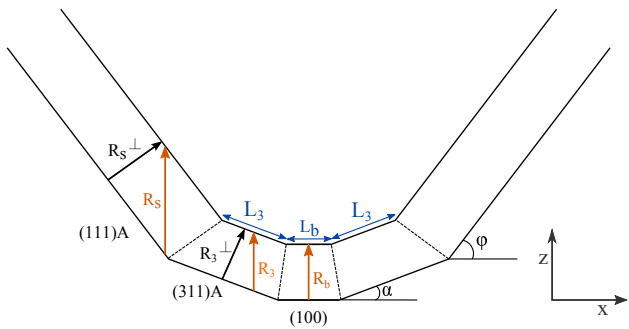}
\caption{(Color online) A two-dimensional section used to model the compositional and morphological evolution within a V-groove.  The labels $b$, $s$, and 3 are used to indicate the base facet, the lateral facets and the intermediate (311)A facets, respectively.}
 \label{fig1}
\end{figure}

\begin{figure}[t]
\includegraphics[width=5cm]{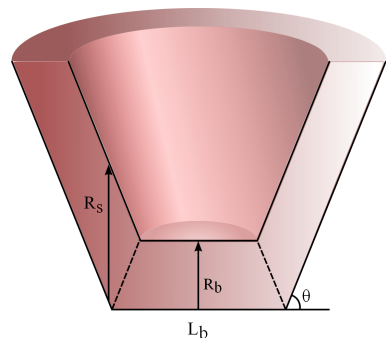}
\caption{(Color online) The template used to model a pyramidal recess, where growth rates, facets and angles are shown.  The labels $b$ and $s$  indicate respectively the base facet and the lateral facets.}
 \label{fig2}
\end{figure}

The solution of (\ref{eq1}) across all facets in the structure requires continuity conditions at each facet boundary for the adatom densities $n_i(x)$ and the corresponding diffusion currents, ${\bf J}_i(x)=-D_i\bm{\nabla}n_i$.  Owing to the translational invariance of V-grooves along their axis, the kinetics will be modelled as the two-dimensional cross-section shown in Fig.~\ref{fig1}.  This assumes that there are no processes along the V-groove that substantially affect the morphological and compositional evolutions. The quality of the fit between experiments and our theory will provide a {\it post hoc} justification of this assumption.  For the growth of QDs in pyramidal recesses, we use the conical template in Fig.~\ref{fig2}, with the circular symmetry about the vertical axis used for simplicity in obtaining an analytic solution of (\ref{eq1}).  
Although the validity of this approximation requires the side facets be much longer than the diffusion lengths of the adatoms, the kinetics exchange mechanisms between the bottom and the side facets are accurately taken into account. When solutions $n_i(x)$ of (\ref{eq1}) are obtained, the growth rate $R_i(x)$ on each facet is expressed as
\begin{equation}
R_i(x)={\Omega_0\over\tau_i}n_i(x)\, ,
\end{equation}
where $\Omega_0$ is the atomic volume.

To calculate the evolution of the facet dimensions during growth, we must solve (\ref{eq1}) coupled to the following equations for the lengths of the facets:
\begin{align}
{dL_b\over dt}&=2\biggl(R_b-{R_3^\perp\over\cos\alpha}\biggr)\cot\alpha\, ,\\
\noalign{\vskip3pt}
{dL_3\over dt}&={\displaystyle{R_3^\perp\over\cos\alpha}-R_b\over\sin\alpha}+{\cos\phi\over\sin(\phi-\alpha)}\biggl({R_3^\perp\over\cos\alpha}-{R_s^\perp\over\cos\phi}\biggr)\, ,
\end{align}
for V-grooves, or coupled with
\begin{equation}
{dL_b\over dt}=2\biggl(R_b-{R_s\over\cos\theta}\biggr)\cot\theta\,
\end{equation}
for pyramidal recesses, where $R_i$ are the average growth rates on each facet, the symbol $\perp$ indicating the component orthogonal to the facet, and $L_i$ are the lengths of the facets comprising the templates. We employ an incremental stationary solution based method to solve the system by choosing a time-step longer than the adatom concentration relaxation time and considering a starting surface profile. Under these assumptions equation (\ref{eq1}) is solved in the stationary regime $(\partial n_i/\partial t=0)$, then  the resulting growth rate on each facet is calculated and the facet dimension variation for each step is found. The iteration of this procedure allows to calculate the time evolution of both surface profile and the relative concentration of the elements in the deposited layers, given by the ratio of the growth rate of each species for each facet.  For long growth times, the vertical growth rates reach a common value, leading to a ``self-limiting'' growth determined by the balance between the diffusion currents and the growth rate anisotropy.

\section{Experiment}
\label{sec3}

\begin{figure}[t]
\includegraphics[width=8.5cm]{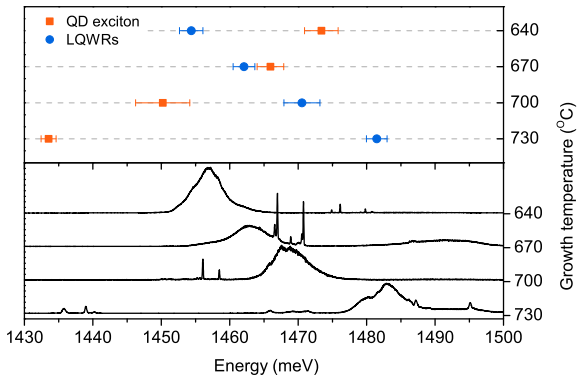}
\caption{(Color online) Photoluminescence spectra of four pyramidal QD samples grown at different temperatures. The temperature dependence of the emission energy of QD and LQWRs obtained from the statistics over a large number of pyramidal QDs is presented in the top graph, while the bottom graph shows four representative spectra.}
 \label{fig3}
\end{figure}

The results obtained from the theoretical model were validated by comparison with published experimental work, as described in the text. In particular, the results of the computations with our model enable the prediction of energy gaps of the different regions of the nanostructures based on their composition, together with a qualitative estimate of the quantum confinement effect, depending on their size. These were then compared with the photo-luminescence spectra obtained performing measurements on the actual samples. 

For experiments carried out for the present study, four pyramidal quantum dot samples (A1-A4) were grown at different growth temperatures (640$^\circ$C, 670$^\circ$C, 700$^\circ$C and 730$^\circ$C) with the aim of exploring the changes in their optical properties\cite{juska14}.  A nominal 0.5-nm-thick In$_{0.25}$Ga$_{0.75}$As layer was grown between two GaAs barriers (the lower being 100 nm thick and the upper 70 nm thick), obtaining a single dot and three lateral wires along the three edges of the pyramidal recess. In Fig.~\ref{fig3} we summarize the photoluminescence spectra of each sample obtained by non-resonant photoexcitation at 8~K relevant to this work, where both the emission originating from the exciton recombination in the QD and from the LQWR are visible, as reported in Ref.~\onlinecite{juska14}. The red-shift of about 30~meV is observed for LQWRs emission, while a blue-shift of about 40~meV is obtained for the QDs as the temperature is decreased. More details relative to the growth of these samples can be found in Ref.~\onlinecite{juska14}. 

The geometrical dependence on the growth temperature and the morphological similarities between V-grooved QWRs and pyramidal LQWRs were analyzed through scanning electron microscopy (SEM) and cross-sectional atomic force microscopy (CS-AFM) of two samples (B1-B2) grown at different temperatures. The structure was the same as for pyramidal QD samples A1-A4 from Ref.~\onlinecite{juska14}, but the growth was interrupted before the In$_{0.25}$Ga$_{0.75}$As layer. A GaAs buffer layer was deposited on top of the GaAs pyramidal recess, then a series of AlGaAs layers of different composition followed by a 100-nm-thick GaAs barrier grown at 640$^\circ$C for sample B1 and 730$^\circ$C for sample B2. Another sample (B2$^\prime$) with the same structure as B2 was grown and capped with another 30 nm Al$_{0.55}$Ga$_{0.45}$As layer to allow a better contrast CS-AFM imaging of the top GaAs layer. The results of our findings will be presented and discussed in the next section.

\section{Results and discussion}

\subsection{Determination of kinetic parameters}

As a first step, our model was employed to simulate the transient growth of In$_{0.12}$Ga$_{0.88}$As V-grooves with GaAs barriers, using as reference model the growth as described in Ref.~\onlinecite{lelarge99}, which reports a transmission electron microscopy (TEM) image clearly showing the morphology and composition of the nanostructure for an alloy with a $12\pm2$\% concentration of In on the sidewalls of the V-groove. Noticeably, the authors report a higher concentration of In on the bottom of the groove measured through electron energy-loss spectroscopy \cite{lelarge99}, specifically of about $22\pm2$\%, probably due to larger diffusion lengths of In adatoms compared to the Ga.

\begin{figure}[b]
\includegraphics[width=8.5cm]{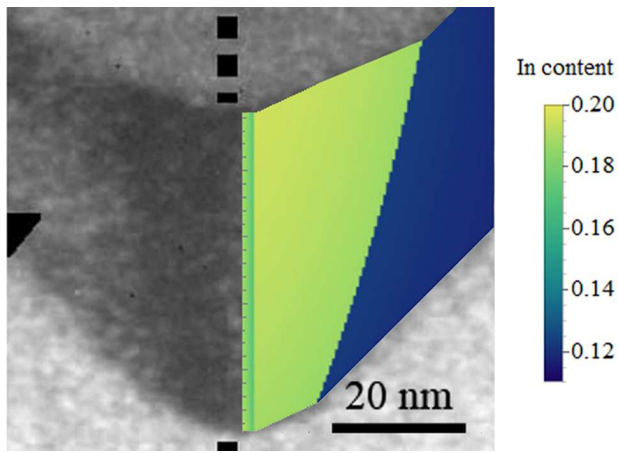}
\caption{(Color online) Result of the transient growth simulation compared to the actual TEM image from Ref.~\onlinecite{lelarge99}, showing good agreement for both dimensional and compositional evolution.}
 \label{fig4}
\end{figure}

After imposing the dimensions and the orientation of the facets composing the initial profile (extrapolated from the reported TEM image, see our Fig.~\ref{fig4}) the transient evolution of the In$_{0.12}$Ga$_{0.88}$As quantum wire on the GaAs V-groove was modelled for a thickness of 45 nm at a ``real'' temperature of 530$^\circ$C (estimated by assuming, since the reactor in Ref.~\onlinecite{lelarge99} is identical to ours, that a similar difference between thermocouple and real growth temperature can be assumed).  The modelling was carried out as in previous work on the (Al)GaAs system, with Ga parameters temperature dependence set identical\cite{dimas13}.  An iterative fitting of the free In kinetic parameters, i.e. the energy barriers for the diffusion and incorporation processes and the effective adatom deposition fluxes (Table~\ref{table1}), produced good agreement between the model and the experimental data for both the morphological and compositional evolution of the InGaAs layer (hence of the In segregation profile). We take this as an indication of the overall validity of the model when applied to InGaAs systems.

The comparison between the two species parameters in Table~\ref{table1} suggests, not unexpectedly, that Indium is a more mobile species, which is consistent with its segregation on the bottom of the V-groove, and is indeed a prerequisite for reproducing the phenomenological findings. 
This is consistent with {\it ab initio} calculations of the diffusion of Ga and In on GaAs(001) surfaces\cite{lepage98,penev01}, which are attributed to the differences  in the cation-As bond strength in the corresponding binary compounds (GaAs,InAs) and the larger ionic radius of indium.  In effect, the potential energy surface is less corrugated for In than for Ga adatoms.

\begin{table}[t]
\begin{ruledtabular}
\renewcommand*{\arraystretch}{1.2}
\caption{Parameter set I for the barriers $E_D$ and $E_\tau$ to diffusion and incorporation, respectively, for In and Ga on the indicated facets. For each facet, once these energy barriers are fixed, the diffusion lengths $\lambda$ are determined ($\lambda=\sqrt{\tau D}$). The parameter $r$ indicates the ratio of the adatom deposition rate on each facet to that relative to the (100) facet. }
\label{table1}
\begin{tabular}{lcccccccc}
&\multicolumn{4}{c}{Indium}&\multicolumn{4}{c}{Gallium}\\
\hline
\multirow{2}{*}{Facet}&$E_D$&$E_\tau$&$\lambda$&$r$&$E_D$&$E_\tau$&$\lambda$&$r$\\
&(eV)&(eV)&(nm)&&(eV)&(eV)&(nm)&\\
\hline
(100)&1.25&0.100&100.7&1.00&1.80&0.114&2.07&1.00\\
(311)A&1.15&0.103&212.5&1.01&1.40&0.128&41.58&1.01\\
(111)A&1.12&0.167&420.1&1.01&1.35&0.159&74.79&1.10\\
\end{tabular}
\end{ruledtabular}
\end{table}

The long In diffusion lengths on the various facets appears to reduce the importance of the decomposition rate anisotropies between facets, as the results of the simulation are largely insensitive to changes in the In ratio for effective deposition fluxes on different facets. For example, varying the ratio of the effective deposition rate on the (111)A facet to that on the (100) facet in the range 0.5 to 2.0 resulted in a change of In concentration of about 1\% only, with a rather small change in the facets dimensions ($<$ 1~nm). 

The nominal relative deposition flux for Indium being 12\%, the segregation level resulting from our simulations is about 20\%, in very good agreement with the experimental values. Furthermore the concentration profile matches that seen in experimental, as evidenced from the TEM color trend in Fig.~\ref{fig4}. Moreover, in the dark-field TEM image from Ref.~\onlinecite{lelarge99}, it is possible to distinguish a vertical region in the center of the V-groove with a lower contrast, which was interpreted as a result of strain or other artifacts.  Our simulations also show a central region above the (100) base facet with a lower In content  (see Fig.~\ref{fig4}, the darker green stripe). Therefore, we suggest that the contrast difference reported in Ref.~\onlinecite{lelarge99} would be simply originating from the morphology of the template and the kinetics during growth, relieving the localized strain factor as the main contribution to segregation effects.

The optimized parameters determined  for V-grooves were then used to simulate the growth in pyramidal recesses\cite{kapon04} at different temperatures.  The three edges (which share the same crystallographic facets and directions as V-grooves) give rise to LQWRs, while the QD forms at the bottom (111)B facet.

\subsection{LQWRs and pyramidal QDs}

In order to apply the model to the LQWR system grown in a pyramidal recess, we make the reasonable assumption that it can be modelled in the same way as the V-groove system in terms of facetting on the bottom of the groove. 
Moreover, the two systems within the pyramidal recess (LQWRs and QD) are assumed independent here in the hypothesis that the region we consider along the LQWR is distant enough from the bottom facet. This is consistent since the dimensions of the pyramid edges are far more extended in length (about 2-3~$\mu m$) than the base facet (30-60 nm).
In these simulations we only consider the last GaAs layer before the In$_{0.25}$Ga$_{0.75}$As layer, and disregard the underlying layer structure under the assumption that the 100-nm-thick GaAs layer has reached its self-limited profile and therefore this profile depends only on the experimental conditions.

\begin{figure}[t]
\includegraphics[width=7cm]{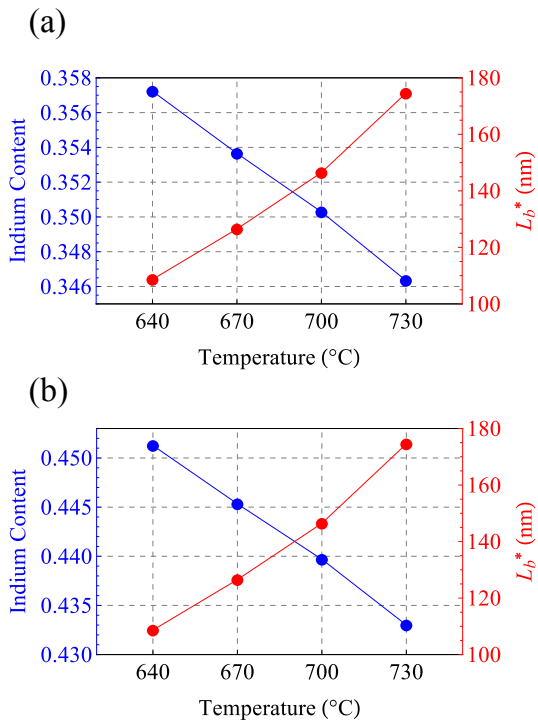}\\

\caption{(Color online) Temperature dependence of In relative content on the bottom facet and of the total length of the base facets $(L_b^\ast=L_b+2L_3)$ resulting from the simulations of LQWRs growth along the edges of a pyramidal recess. Results with (a) parameter set I (Table~\ref{table1}),  and (b) parameter set 2 (Table~\ref{table2}).}
 \label{fig5}
\end{figure}

Using the parameters set in Table~\ref{table1}, our model was applied to the LQWRs in the pyramidal recesses by simulating the growth of 2 nm of In$_{0.25}$Ga$_{0.75}$As over the GaAs self-limited profile (whose dimensions were calculated following Ref.\onlinecite{dimas13}) at the four different growth temperatures of the samples (A1-A4) described in Sec.~\ref{sec3}. Note that the nominal thickness of the dot layer is 0.5 nm. However, in our growth regime, an overall increase of the vertical growth rate at the bottom of the template occurs, leading to an increased vertical thickness of the deposited layer on the base facet. The chosen value for the simulations (2 nm) is a typical value, as estimated in previous theoretical works\cite{healy10}. As a result of each simulation, both composition and lateral dimensions were found not to vary significantly over the vertical thickness of 2 nm during the transient evolution, so we show only an average value in Fig.~\ref{fig5}. The simulations indeed evidence an increase in In segregation at the bottom of the LQWR as the temperature is decreased, which in terms of the energy gap, leads to a decrease of about 11.7~meV at 8~K, without considering confinement effects.  This is in line with the experimentally reported red-shift discussed earlier, even if the experimentally measured value is bigger ($\sim$30~meV).

To predict the emission-energy-dependence on the growth temperature of the nanostructure we must take into account the change in the quantum confinement effect derived from the variation of the  dimension. Nevertheless, in the case of these LQWRs, we may be able to assume that this is  a small effect, since the total lateral dimensions $(L_b^\ast=L_b+2L_3)$ are far larger (from 108 nm to 175 nm, depending on the growth temperature) than the Bohr radius of the exciton (which can be estimated as about 20-30 nm), and only the vertical dimension should be affected. The morphological/geometrical similarity between the two systems (V-grooved quantum wires and LQWRs) will be discussed in the next section, in which a series of deviations from the ideal situation will be presented.

\begin{table}[t]
\begin{ruledtabular}
\renewcommand*{\arraystretch}{1.2}
\caption{Parameter set II for the barriers $E_D$ and $E_\tau$ to diffusion and incorporation, respectively,for In and Ga on the indicated facets.  The parameter $r$ indicates the decomposition rate anisotropy relative to (100).}
\label{table2}
\begin{tabular}{lccccccc}
&\multicolumn{3}{c}{Indium}&\multicolumn{3}{c}{Ga}\\
\hline
\multirow{2}{*}{Facet}&$E_D$&$E_\tau$&$r$&$E_D$&$E_\tau$&$r$\\
&(eV)&(eV)&&(eV)&(eV)&\\
\hline
(100)&1.27&0.090&1.00&1.85&0.122&1.00\\
(311)A&1.22&0.095&1.01&1.55&0.141&1.01\\
(111)A&1.20&0.187&1.01&1.25&0.159&1.10\\
\end{tabular}
\end{ruledtabular}
\end{table}

Keeping in mind the non-ideal nature of the actual samples, in order to verify that the segregation temperature dependence could have a major role in the red-shift, a further study was done.  Another iterative procedure was carried out to determine a set of kinetic parameters for both Ga and In that would result in an even more pronounced shift of the spectrum along with the temperature without deviating too much from the parameters obtained from the previous fit. The resulting set of parameters (Table~\ref{table2}) resulted in the temperature dependence shown in Fig.~\ref{fig5}(b), corresponding to an energy gap decrease of about 21.1~meV. Although this result was not obtained directly from the fitting of experimental data, it shows that the hypothesis of segregation temperature dependence could be a valid explanation for the observed red-shift and indeed a compatible physical process in this system.

The next step was to apply the model to the growth of the InGaAs QD using the three-dimensional (3D) conical representation in Fig.~\ref{fig2} to verify that the same temperature dependent mechanism does not significantly affect the QD (as expected from the blue shift reported, which was tentatively attributed in Ref.~\onlinecite{juska14} to a change in the self-limited profile, and not to a change in In segregation). The kinetic parameters for the lateral (111)A facet were chosen to be equal to those relative to the same facet in the LQWR growth model. For the base (111)B facet, however, no detailed experimental data was available to fit the growth of In$_{0.25}$Ga$_{0.75}$As on GaAs. A set of kinetic parameters was chosen that resulted in a segregation of In of about 4\% on the bottom facet at a specific growth temperature, as suggested by other theoretical studies on pyramidal QD optical properties\cite{healy10}.

\begin{table}[t]
\begin{ruledtabular}
\renewcommand*{\arraystretch}{1.2}
\caption{Parameters for the (111)B facet optimized for parameter sets I (Table~\ref{table1}) and II (Table~\ref{table2}).}
\label{table3}
\begin{tabular}{ccccccccc}
&\multicolumn{4}{c}{Indium}&\multicolumn{4}{c}{Ga}\\
\hline
Optimized&$E_D$&$E_\tau$&$\nu_0$&$r$&$E_D$&$E_\tau$&$\nu_0$&$r$\\
for set&(eV)&(eV)&(s\textsuperscript{-1})&&(eV)&(eV)&(s\textsuperscript{-1})&\\
\hline
1&1.45&0.059&5.81&1.01&1.52&0.031&4.13&1.10\\
2&1.50&0.066&5.71&1.01&1.55&0.020&4.00&1.10\\
\end{tabular}
\end{ruledtabular}
\end{table}

\begin{figure}[b]
\includegraphics[width=7cm]{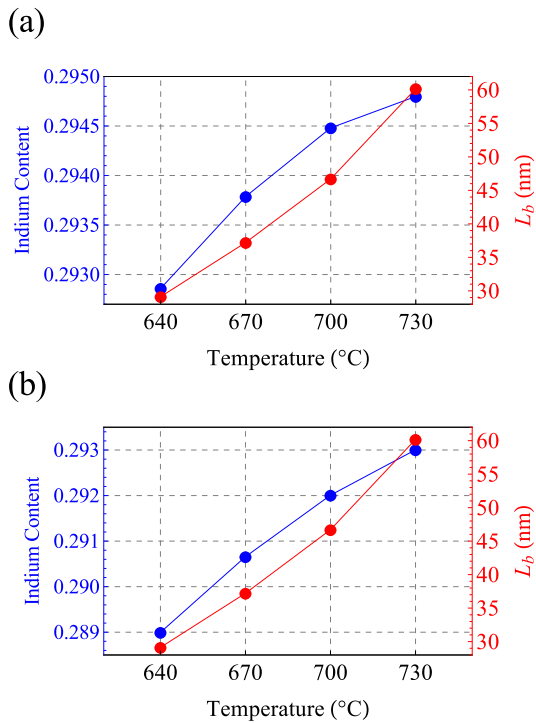}
\caption{(Color online) Temperature dependence of the length and In relative concentration on the base facet resulting from the simulations of the QD growth on the bottom of a pyramidal recess. Results obtained with (a) parameters set I (Table~\ref{table1}), and (b) parameters set II (Table~\ref{table2}).}
 \label{fig6}
\end{figure}

As pointed out elsewhere\cite{dimas12}, the fitting of the parameters for the 3D case requires the optimization of the exponential prefactors $\nu_0$ of the adatom lifetimes in order to get a consistent result. This was done for parameter sets 1 and 2 and for both Ga and In. For each set we found parameters for the (111)B bottom facet (Table~\ref{table3}) that resulted in around 4\% segregation on that facet, interestingly, without presenting any significant increase of In concentration as the temperature was increased (Fig.~\ref{fig6}), rather differently from the LQWR case. In this case the lateral dimension of the QDs are comparable to the exciton Bohr radius, leading to quantum-confinement influence on the emission.  In particular the decrease in temperature implies a decrease in the lateral dimensions of the QD, and therefore to a blue-shift of the emission, as observed experimentally. Nevertheless, we caution the reader that more theoretical calculations need to be done in order to evaluate completely the origin of this effect in our case, taking into account the particular geometry of the QD.

\subsection{Morphology of pyramidal recesses}

\begin{figure}[t]
\includegraphics[width=8cm]{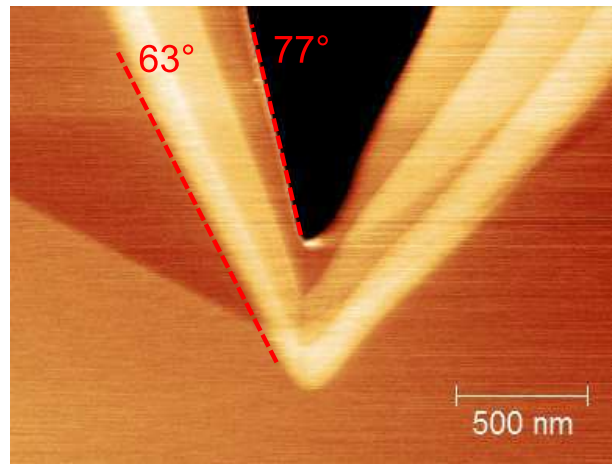}\\
\caption{(Color online) CS-AFM image of the vertical section of a pyramidal recess from sample B2$^\prime$ showing the layer sequence and the evolution of the orientation of the lateral (111)A vicinal facets.}
 \label{fig7}
\end{figure}

To gain  insight into the quantitative discrepancies between the experimental theoretical results, SEM and CS-AFM imaging was performed on representative samples, which revealed that the actual shape of a pyramidal recess is made of a more complex facetting than the simplified profile assumed in our and previous models. Samples B1 and B2, described in Sec.~\ref{sec3},  were cleaved along the (110) direction in order to image the cross-section of both the LQWRs and QD positions, with the nanostructure which could be imaged depending on the point where the cleavage was actually done.

CS-AFM analysis performed on samples B1 and B2 showed that the lateral facet (111)A orientation evolves during the growth of the different layers (this is not in itself a new observation, but we recall it here for completeness). Therefore, the final GaAs facet is not a pure (111)A, but a {\it vicinal} facet. The reason for this phenomenon is not clear, but we notice that it holds similarity to what happens in the case of V-groove quantum wires\cite{dimas13}. Considering, for example, the sample shown in Fig.~\ref{fig7}, the angle between the vicinal facet (111)A and the base facet (111)B is 77$^\circ$, and from basic trigonometry, the resulting angle between the lateral facet (111)A and the edge base facet (100) is about 33$^\circ$, which significantly differs from the 45$^\circ$ angle that characterizes the V-groove. This suggests that the kinetic parameters for both Ga and In should be optimized for the particular vicinal facet, and could be one of the reasons for specific deviations between theory and experiment.

SEM imaging was also performed in both top-view and tilted view (in order to distinguish the cross-section of the pyramid). Particular care was taken in order to make sure that (or searching for regions where) the cleaving line passed through a precise point of the pyramidal recess to enable the edge of the pyramid and its center cross-section to be distinguished. Unless specified, the following considerations are valid for both samples B1 and B2, which showed very similar qualitative characteristics.

\begin{figure*}[t]
\includegraphics[width=16.2cm]{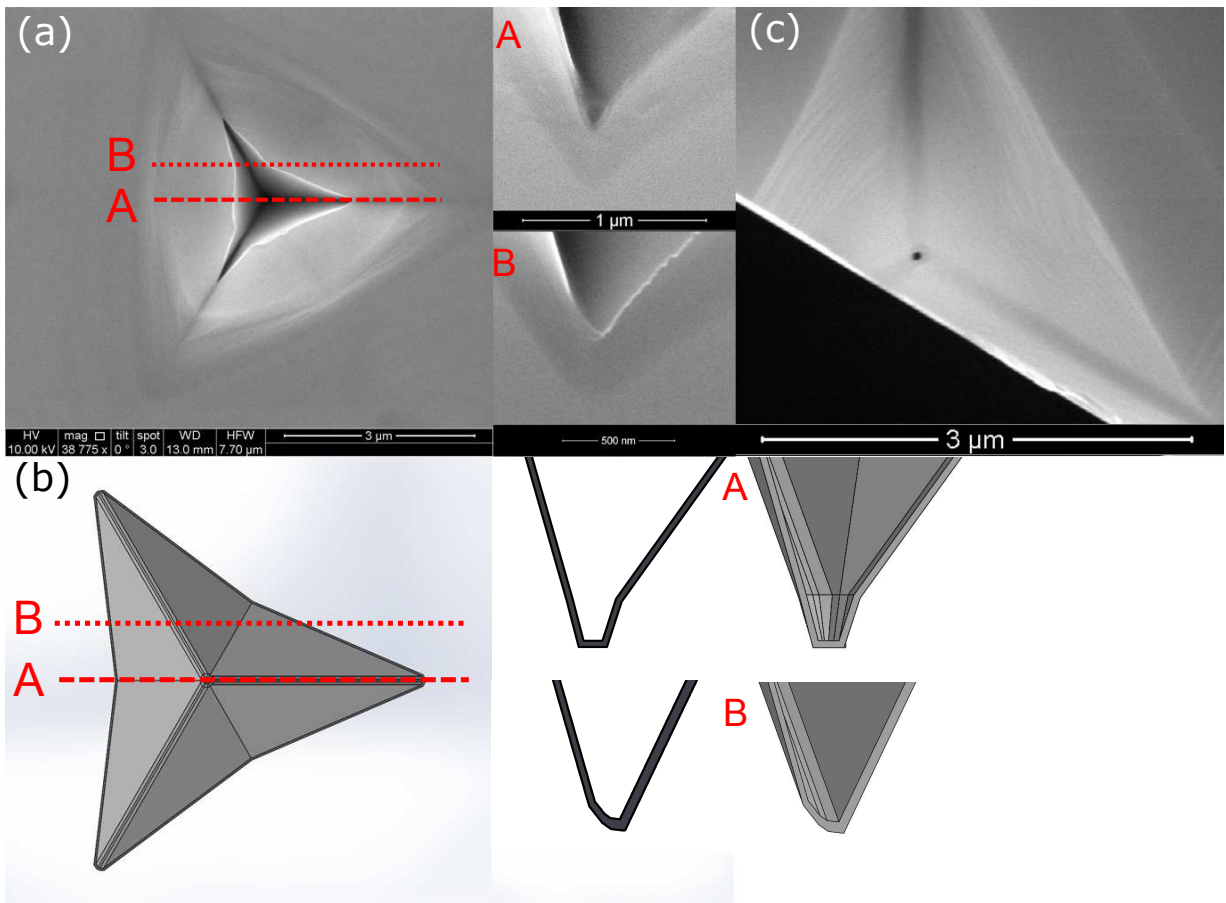}\\
\caption{(Color online) (a) Top-view SEM image of a pyramidal recess from sample B1, and cross-sectional views of recesses cleaved along different sections (cross-section A is relative to a cleavage cutting through the center of the pyramid, while cross section B to one along an axis cutting through one of the LQWRs).  (b) 3D model of a pyramidal recess, reconstructed considering the experimental results of our analysis and qualitative comparison of the obtained cross-sections. A and B cross-sections correspond to the same labeled experimental cross-sections, and for each both the 3D cross-section (on the right) and the upper surface outline (on the left) are shown. (c) top-view SEM image of a pyramidal recess from sample B2 after cleavage; this particular shows the different depths of the edges and of the bottom.}
 \label{fig8}
\end{figure*}

The top-view images in Fig.~\ref{fig8} shows that each of the three lateral facets of the pyramid is formed by two vicinal facets, creating a sort of hexagonal inward-shaped top outline (Fig.~\ref{fig8}(a)). This effect is more pronounced in the lower-temperature sample (B1), and less evident for high temperature growth (B2). The top-view images also show a dark region corresponding to the edges and bottom of the pyramid, denoting deeper regions of the recess (we should remark that it not so evident at first, but this becomes clearer after a number of these analyses are performed). In particular the center of the recess was found to be the deepest feature, delimited by a quasi-circular outline (Fig.~\ref{fig8}(c)), smaller in diameter than the lateral broadening of the wires. This suggests that the bottom facetting of the pyramid cannot be simply related to the LQWRs and that it is likely more complicated than the facetting structure assumed in our model.  For example, a steeper (vicinal) facet between the lateral vicinal (111)A and the bottom (111)B facets, resulting from the assembly of the three pyramid edges joined at the center of the recess, could lead to a more pronounced dip. 

Using the foregoing observations and hypothesizing that that the facetting at the edges is similar to that of a V-groove, taking all into consideration, we built a qualitative graphical 3D model of the pyramidal recess, from which we obtained the corresponding cross-sectional view in different points of the recess (Fig.~\ref{fig8}(a,b)). A comparison between the experimental and hypothesized cross-section shows that our 3D model is compatible with the observed experimental cross sectional morphology.  Moreover, recent Monte Carlo growth simulations from Ref.~\onlinecite{surrente13} suggested a similar behavior of the facetting on the bottom recess of (low temperature grown) small pitch pyramidal recesses, and are therefore in agreement with our findings. Obviously more microscopy work (TEM) will be needed, both to clarify the exact morphological facetting at the bottom of the recesses, and to understand the exact implications in terms of optical properties.

\section{Conclusions}

In this work the growth model for MOVPE on patterned substrates we previously presented was extended to the growth of InGaAs nanostructures. The kinetic parameters resulting from an empirical fitting procedure to the experimental data of a V-grooved quantum wire from the literature produced good agreement between the simulated and the actual growth result in terms of morphology and composition, reproducing the segregation profile of In on the bottom of the V-groove. Therefore, on one hand, this result can be considered a validation of the model and, on the other hand, enables simulations of the growth of other nanostructures to be carried out. Here, through simulations at different growth temperatures for pyramidal QDs and LQWRs a new explanation for the unexpected reported behavior of the LQWRs emission in pyramidal QDs was proposed. 

The model will be employed in future work in order to guide the MOVPE growth parameters and provide a better control over nanostructure formation. As a first step, the analysis carried out through SEM and CS-AFM to understand the actual geometry of a LQWR and its compatibility with the V-groove picture revealed a complicated facetting on the bottom of the pyramid. Our findings will be important for the scientific community for correlating/describing pyramidal QD optical properties and will be further investigated to optimize the theoretical model and obtain more accurate simulations.

\section*{Acknowledgements}

This research was partly enabled by the Irish Higher Education Authority Program for Research in Third Level Institutions (2007-2011) via the INSPIRE programme, by Science Foundation Ireland under grants 10/IN.1/I3000. We thank K.~Thomas for his support with the MOVPE system.

\end{document}